\documentclass[12pt]{article}
\usepackage{graphicx}
\def\beq{\begin{equation}}
\def\eeq{\end{equation}}
\def\bea{\begin{eqnarray}}
\def\eea{\end{eqnarray}}
\def\nn{\nonumber}
\def\roughly#1{\mathrel{\raise.3ex\hbox
{$#1$\kern-.75em\lower1ex\hbox{$\sim$}}}}

\def\ks{K_S}
\def\kbar{{\bar K}^0}
\def\bd{B_d^0}
\def\bs{B_s^0}
\def\bdbar{{\bar B}_d^0}
\def\bsbar{{\bar B}_s^0}
\def\btod{{\bar b} \to {\bar d}}
\def\btos{{\bar b} \to {\bar s}}
\def\Bsdecay{\bsbar\to J/\psi \phi}
\def\bskk{\bs\to K^{(*)0} {\bar K}^{(*)0}}
\def\bdkk{\bd\to K^{(*)0} {\bar K}^{(*)0}}

\begin{document}

\begin{flushright}
UdeM-GPP-TH-12-206 \\
UMISS-HEP-2012-02\\
\end{flushright}

\begin{center}
\bigskip
{\Large \bf \boldmath Measuring $\beta_s$ with $\bskk$ -- a Reappraisal} \\
\bigskip
\bigskip
{\large 
Bhubanjyoti Bhattacharya $^{a,}$\footnote{bhujyo@lps.umontreal.ca},
Alakabha Datta $^{b,}$\footnote{datta@phy.olemiss.edu}, \\
Maxime Imbeault $^{c,}$\footnote{mimbeault@cegep-st-laurent.qc.ca}
and David London $^{a,}$\footnote{london@lps.umontreal.ca}
}
\end{center}

\begin{flushleft}
~~~~~~~~~~~$a$: {\it Physique des Particules, Universit\'e
de Montr\'eal,}\\
~~~~~~~~~~~~~~~{\it C.P. 6128, succ. centre-ville, Montr\'eal, QC,
Canada H3C 3J7}\\
~~~~~~~~~~~$b$: {\it Department of Physics and Astronomy, 108 Lewis Hall, }\\ 
~~~~~~~~~~~~~~~{\it University of Mississippi, Oxford, MS 38677-1848, USA}\\
~~~~~~~~~~~$c$: {\it D\'epartement de physique, C\'egep de Saint-Laurent,}\\
~~~~~~~~~~~~~~~{\it 625, avenue Sainte-Croix, Montr\'eal, QC, Canada H4L 3X7 }
\end{flushleft}

\begin{center}
\bigskip (\today)
\vskip0.5cm {\Large Abstract\\} \vskip3truemm
\parbox[t]{\textwidth}{The $\bs$-$\bsbar$ mixing phase, $\beta_s$, can
  be extracted from $\bskk$, but there is a theoretical error if the
  second amplitude, $V_{ub}^* V_{us} P'_{uc}$, is non-negligible.
  Ciuchini, Pierini and Silvestrini (CPS) have suggested measuring
  $P_{uc}$ in $\bdkk$, and relating it to $P'_{uc}$ using SU(3). For
  their choice of the direct and indirect CP asymmetries in $\bdkk$,
  they find that the error on $\beta_s$ is very small, even allowing
  for 100\% SU(3) breaking. In this paper, we re-examine the CPS
  method, allowing for a large range of the $\bdkk$ observables. We
  find that the theoretical error in the extraction of $\beta_s$ can
  be quite large, up to $18^\circ$. This problem can be ameliorated if
  the value of SU(3) breaking were known, and we discuss different
  ways, both experimental and theoretical, of determining this
  quantity.}

\end{center}

\thispagestyle{empty}
\newpage
\setcounter{page}{1}
\baselineskip=14pt

\section{Introduction}

In the standard model (SM), the weak phase of $\bs$-$\bsbar$ mixing,
$\beta_s$, is $\approx 0$. Thus, if its value is measured to be
nonzero, this is a clear sign of new physics (NP). Indeed, experiments
have already started measuring $\beta_s$ in $\Bsdecay$. The results of
the CDF \cite{CDF} and D\O\ \cite{D0} collaborations hint at NP, but
the errors are very large. On the other hand, the LHCb collaboration
\cite{LHCb_betas} finds a central value for $\beta_s$ which is
consistent with zero: $\beta_s = (-0.03 \pm 2.89~({\rm stat}) \pm
0.77~({\rm syst}))^\circ$, implying that, if NP is present in
$\bs$-$\bsbar$ mixing, its effect is small.

In the $\bd$ system, the phase of $\bd$-$\bdbar$ mixing, $\beta$, was
first measured in the ``golden mode'' $\bd\to J/\psi \ks$, and
subsequently in many other modes such as $\btos$ penguin decays
(e.g.\ $\bd\to \phi \ks$), ${\bar b} \to {\bar c}c{\bar d}$ decays
(e.g.\ $\bd\to J/\psi \pi^0$), etc. In the same vein, it is important
to measure $\beta_s$ in many different decay modes.

One process which is potentially a good candidate for measuring
$\beta_s$ is the pure $\btos$ penguin decay $\bskk$. Its amplitude can
be written
\beq
{\cal A}_s = V_{ub}^* V_{us} P'_{uc} + V_{tb}^* V_{ts} P'_{tc} ~.
\eeq
Now, we know that $|V_{ub}^* V_{us}|$ and $|V_{tb}^* V_{ts}|$ are
$O(\lambda^4)$ and $O(\lambda^2)$, respectively, where $\lambda=0.23$
is the sine of the Cabibbo angle. This suggests that the $V_{ub}^*
V_{us} P'_{uc}$ term is possibly negligible compared to $V_{tb}^*
V_{ts} P'_{tc}$. If this is justified, then there is essentially only
one decay amplitude, and $\beta_s$ can be cleanly extracted from the
indirect CP asymmetry in $\bskk$.

The difficulty is that it is not completely clear whether $V_{ub}^*
V_{us} P'_{uc}$ is, in fact, negligible. This term has a different
weak phase than that of $V_{tb}^* V_{ts} P'_{tc}$, so that its
inclusion will ``pollute'' the extraction of $\beta_s$. That is, if it
contributes significantly to the amplitude, the value of $\beta_s$
measured in $\bskk$ will deviate from the true value of $\beta_s$, and
this theoretical error is directly related to the relative size of the
two terms.

This issue has been examined by Ciuchini, Pierini and Silvestrini
(CPS) in Ref.~\cite{CPS}. In order to get a handle on the size of
$P'_{uc}$, CPS proceeded as follows. They considered the
U-spin-conjugate decay, $\bdkk$, focusing specifically on $\bd \to
K^{*0} {\bar K}^{*0}$.  This is a pure $\btod$ penguin decay, whose
amplitude is
\beq
{\cal A}_d = V_{ub}^* V_{ud} P_{uc} + V_{tb}^* V_{td} P_{tc} ~.
\eeq
If one takes the values for the CKM matrix elements, including the
weak phases, from independent measurements, then this amplitude
depends only on three unknown parameters: the magnitudes of $P_{uc}$
and $P_{tc}$, and their relative strong phase. But there are three
experimental measurements one can make of this decay -- the branching
ratio, the direct CP asymmetry, and the indirect (mixing-induced) CP
asymmetry. It is therefore possible to solve for all the unknown
parameters. In particular, one can obtain $|P_{uc}|$. This quantity
can be related to $|P'_{uc}|$ by an SU(3)-breaking factor. Now, in
2007, when Ref.~\cite{CPS} was written, there were no experimental
measurements of $B_{d,s}^0 \to K^{*0} {\bar K}^{*0}$. Instead, CPS
assumed values for these measurements, inspired by QCD factorization
(QCDf) \cite{BBNS}. They found that, even allowing for 100\% SU(3)
breaking, the value of $|P'_{uc}|$ is such that the error on $\beta_s$
due to the inclusion of a nonzero $V_{ub}^* V_{us} P'_{uc}$ term is
less than $1^\circ$. This inspired CPS to dub $\bskk$ the {\it golden
  channel} for measuring $\beta_s$.

In this paper, we re-examine the method of CPS. In particular, we want
to establish to what extent CPS's conclusion is dependent on the
values chosen for the $\bdkk$ experimental observables. As we will
see, the CPS result holds for a significant subset of the input
values. However, it also fails for other choices of the inputs -- the
error on $\beta_s$ due to the presence of the $V_{ub}^* V_{us}
P'_{uc}$ term can be as large as $18^\circ$. It is therefore not
correct to say that the $V_{ub}^* V_{us} P'_{uc}$ term has little
effect, i.e.\ that $\beta_s$ can always be measured cleanly in
$\bskk$. On the other hand, it is true that $|P_{uc}|$ can be
extracted from $\bdkk$. This can then be used to obtain information
about $|P'_{uc}|$ if the SU(3)-breaking factor were known reasonably
accurately. We discuss different ways, both experimental and
theoretical, of learning about the size of the SU(3) breaking.

In Sec.~2, we examine $B^0_{d,s} \to K^{(*)0} {\bar K}^{(*)0}$, and
show how the $\bd$ decay can be used to obtain information about the
$\bs$ decay. We allow for all values of the observables in the $\bd$
decay, and compute the theoretical error on $\beta_s$, allowing for
100\% SU(3) breaking. It turns out that this error can be
substantial. In Sec.~3, we discuss ways, both experimental and
theoretical, of determining the SU(3) breaking. If this breaking is
known with reasonable accuracy, this greatly reduces the theoretical
error on $\beta_s$, and allows this mixing quantity to be extracted
from $B^0_{d,s} \to K^{(*)0} {\bar K}^{(*)0}$ decays. We conclude in
Sec.~4.

\section{\boldmath $B^0_{d,s} \to K^{(*)0} {\bar K}^{(*)0}$}

\subsection{\boldmath $\bskk$}
\label{bskkSec}

$\bskk$ is a pure $\btos$ penguin decay. That is, its amplitude
receives contributions only from gluonic and electroweak penguin (EWP)
diagrams.  There are three contributing amplitudes, one for each of
the internal quarks $u$, $c$ and $t$ (the EWP diagram contributes only
to $P'_t$):
\bea
{\cal A}_s &=& \lambda^{(s)}_u P'_u + \lambda^{(s)}_c P'_c + \lambda^{(s)}_t P'_t \nn\\
         &=& |\lambda^{(s)}_u| e^{i\gamma} P'_{uc} - |\lambda^{(s)}_t| P'_{tc} ~,
\label{Bsamp}
\eea
where $\lambda^{(q')}_q \equiv V_{qb}^* V_{qq'}$.  (As this is a
$\btos$ transition, the diagrams are written with primes.) In the
second line, we have used the unitarity of the
Cabibbo-Kobayashi-Maskawa (CKM) matrix ($\lambda^{(s)}_u +
\lambda^{(s)}_c + \lambda^{(s)}_t = 0$) to eliminate the $c$-quark
contribution: $P'_{uc} \equiv P'_u - P'_c$, $P'_{tc} \equiv P'_t -
P'_c$. Also, above we have explicitly written the weak-phase
dependence (including the minus sign from $V_{ts}$ in
$\lambda^{(s)}_t$), while $P'_{uc}$ and $P'_{tc}$ contain strong
phases. (The phase information in the CKM matrix is conventionally
parametrized in terms of the unitarity triangle, in which the interior
(CP-violating) angles are known as $\alpha$, $\beta$ and $\gamma$
\cite{pdg}.)  The amplitude ${\bar {\cal A}}_s$ describing the
CP-conjugate decay $\bsbar\to K^{(*)0} {\bar K}^{(*)0}$ can be
obtained from the above by changing the signs of the weak phases (in
this case, $\gamma$).

There are three measurements which can be made of $\bskk$: the
branching ratio, and the direct and indirect CP-violating asymmetries.
These yield the three observables
\bea
\label{'observables}
X' & \equiv & \frac{1}{2} \left( |{\cal A}_s|^2 + |{\bar{\cal A}}_s|^2 \right) ~, \nn \\
Y' & \equiv & \frac{1}{2} \left( |{\cal A}_s|^2 - |{\bar{\cal A}}_s|^2 \right) ~, \nn\\
Z'_I & \equiv & {\rm Im}\left( e^{-2i \beta_s} {\cal A}^*_s {\bar {\cal A}}_s \right) ~.
\eea
Assuming one takes the values for $|\lambda^{(s)}_u|$,
$|\lambda^{(s)}_t|$ and $\gamma$ from independent measurements, ${\cal
  A}_s$ then depends only on the magnitudes of $P'_{uc}$ and
$P'_{tc}$, and their relative strong phase $\delta'$. With $\beta_s$,
this makes a total of four unknown parameters. These cannot be
determined from only three observables -- additional input is needed.

Note that, if $\lambda^{(s)}_u P'_{uc}$ were negligible, we would only
have two unknowns -- $|P'_{tc}|$ and $\beta_s$. These could be
determined from the measurements of $X'$ and $Z'_I$ ($Y'$ would
vanish). This demonstrates that if one extracts $\beta_s$ from $Z'_I$
assuming that $\lambda^{(s)}_u P'_{uc}$ is negligible, and it is not,
then one will obtain an incorrect value for $\beta_s$. The size of
this error is directly related to the size of $\lambda^{(s)}_u
P'_{uc}$. Here, the possibility of an error is particularly
important. Since $\beta_s \approx 0$ in the SM, a nonzero measured
value of $\beta_s$ would indicate NP\footnote{Note that, if there is
  an indication of NP, we will know that it is in $\btos$
  transitions. However, we will not know if $\bs$-$\bsbar$ mixing
  and/or the $\btos$ penguin amplitude is affected.}. It is therefore
crucial to have this theoretical uncertainty under control.

\subsection{\boldmath $\bdkk$}

In order to deal with the $P'_{uc}$ problem in $\bskk$, in
Ref.~\cite{CPS}, CPS use its U-spin-conjugate decay $\bdkk$ . This is
a pure $\btod$ penguin decay, whose amplitude can be written
\beq
{\cal A}_d = |\lambda^{(d)}_u| e^{i\gamma} P_{uc} + |\lambda^{(d)}_t| e^{-i\beta} P_{tc} ~.
\eeq
As with ${\cal A}_s$, we take the values for the magnitudes and weak
phases of the CKM matrix elements from independent measurements. This
leaves three unknown parameters in ${\cal A}_d$: the magnitudes of
$P_{uc}$ and $P_{tc}$, and their relative strong phase $\delta$.  And,
as with $\bskk$, there are three measurements which can be made of
$\bdkk$: the branching ratio, and the direct and indirect CP-violating
asymmetries. Given an equal number of observables and unknowns, we can
solve for $|P_{uc}|$, $|P_{tc}|$ and $\delta$.

The key point is that $|P_{uc}|$ and $|P'_{uc}|$ are equal under
flavor SU(3) symmetry. Thus, given a value for $|P_{uc}|$ and a value
(or range) for the SU(3)-breaking factor, one obtains the value (or
range) of $|P'_{uc}|$. With this, one can extract the true value (or
range) of $\beta_s$ from the $\bskk$ experimental data.

Now, CPS focused mainly on the decays $B_{d,s}^0 \to K^{*0} {\bar
  K}^{*0}$. As mentioned in the introduction, there were no
experimental measurements of these decays when their paper was
written, so it was necessary to assume experimental values in order to
extract $|P_{uc}|$. CPS chose values roughly based on the QCDf
calculation of Ref.~\cite{BRY}. They found that the value of
$|P_{uc}|$ is such that, even allowing for 100\% SU(3) breaking,
$|\lambda^{(s)}_u P'_{uc}|$ is indeed small. The upshot is that the
theoretical uncertainty in the extraction of $\beta_s$ is less than
$1^\circ$.

There are several reasons not to take this result at face value.
First, although QCDf has been very successful at describing the
$B$-decay data, it is still a model. Indeed, the predictions and
explanations of other models of QCD -- perturbative QCD \cite{pQCD}
(pQCD) and SCET \cite{SCET}, for example -- do not always agree with
those of QCDf. Second, QCDf assumes that factorization holds to
leading order for all $B$ decays. However, $B^0_{d,s} \to K^{(*)0}
{\bar K}^{(*)0}$ are penguin decays, and it has been argued that
non-factorizable effects are important for such decays. It may be that
sub-leading effects in QCDf are, in fact, important for $\bd \to
K^{*0} {\bar K}^{*0}$. We therefore re-examine the CPS method taking a
more model-independent approach.

\subsection{\boldmath Theoretical Uncertainty on $\beta_s$}

In this subsection, we generalize the CPS method. First, we consider
all final states in $B^0_{d,s} \to K^{(*)0} {\bar K}^{(*)0}$. Second,
we scan over a large range of experimental input values.

We proceed as follows. The three experimental measurements of the
$\bd$ decay correspond to the three observables
\bea
\label{XYZdefs}
X & \equiv & \frac{1}{2} \left( |{\cal A}_d|^2 + |{\bar{\cal A}}_d|^2 \right) \nn\\
&& \hskip0.4truecm 
=~|\lambda^{(d)}_u|^2 |P_{uc}|^2 + |\lambda^{(d)}_t|^2 |P_{tc}|^2 - 2 |\lambda^{(d)}_u||\lambda^{(d)}_t||P_{uc}| |P_{tc}| \cos\delta \cos\alpha ~, \nn \\
Y & \equiv & \frac{1}{2} \left( |{\cal A}_d|^2 - |{\bar{\cal A}}_d|^2 \right) =
- 2 |\lambda^{(d)}_u||\lambda^{(d)}_t||P_{uc}| |P_{tc}| \sin\delta \sin\alpha ~, \\
Z_I & \equiv & {\rm Im}\left( e^{-2i \beta} {\cal A}_d^* {\bar {\cal A}}_d \right)
= |\lambda^{(d)}_u|^2 |P_{uc}|^2 \sin 2\alpha - 2 |\lambda^{(d)}_u||\lambda^{(d)}_t||P_{uc}| |P_{tc}| \cos\delta \sin\alpha ~. \nn
\eea
It is useful to define a fourth observable:
\bea
\label{ZRdef}
Z_R & \equiv & {\rm Re}\left( e^{-2i \beta} {\cal A}_d^* {\bar {\cal A}}_d \right) \\
&& \hskip0.4truecm 
=~|\lambda^{(d)}_u|^2 |P_{uc}|^2 \cos 2\alpha + |\lambda^{(d)}_t|^2 |P_{tc}|^2 - 2 |\lambda^{(d)}_u||\lambda^{(d)}_t| |P_{uc}| |P_{tc}| \cos\delta \cos\alpha ~. \nn
\eea
The quantity $Z_R$ is not independent of the other three observables:
\beq
Z_R^2 = X^2 - Y^2 - Z_I^2 ~.
\label{eq:ZR}
\eeq
Thus, one can obtain $Z_R$ from measurements of $X$, $Y$ and $Z_I$, up
to a sign ambiguity.

$X$, $Y$ and $Z_I$ are related to the branching ratio
($\mathcal{B}_d$), the direct CP asymmetry ($C_d$) and the indirect CP
asymmetry ($S_d$) of $\bdkk$ as follows:
\beq
X = \kappa_d \mathcal{B}_d ~~,~~~~ 
Y = \kappa_d \mathcal{B}_d C_d ~~,~~~~
Z_I = \kappa_d \mathcal{B}_d S_d ~,
\eeq
where
\beq
\kappa_d = \frac{8 \pi m_{B_d}^2}{\tau_d p_c} ~.
\eeq
In the above, $m_{B_d}$ and $\tau_d$ are the mass and the lifetime of
the decaying $\bd$ meson, respectively, and $p_c$ is the momentum of
the final-state mesons in the rest frame of the $\bd$. 

{}From Eqs.~(\ref{XYZdefs}) and (\ref{ZRdef}), the quantity $|P_{uc}|$
can then be written in terms of the observables as
\beq
|P_{uc}|^2 = \frac{1}{|\lambda^{(d)}_u|^2} \, \frac{Z_R - X}{\cos 2\alpha - 1}
= \frac{\kappa_d \mathcal{B}_d}{|\lambda^{(d)}_u|^2} \, \frac{\pm \sqrt{1-C_d^2-S_d^2} - 1}{\cos 2\alpha - 1} ~.
\label{eq:Puc2}
\eeq
The value of $\alpha$ is not known exactly, but we know from
independent measurements that it is approximately $90^\circ$. In what
follows, we fix $\alpha$ to $90^\circ$ for simplicity.  Note that any
deviation of $\alpha$ from this value decreases the denominator in
Eq.~(\ref{eq:Puc2}), and thus makes $|P_{uc}|$ larger. The above
expression allows us to calculate $|P_{uc}|$ for a given set of
observables.

On the whole, the decays $\bdkk$ have not yet been measured. One
exception is $\bd \to K_S K_S$. From BaBar \cite{Aubert:2006gm}, we
have
\beq
\mathcal{B}_d = (1.08 \pm 0.28 \pm 0.11) \times 10^{-6} ~~,~~~~
S_d = -1.28^{+0.80+0.11}_{-0.73-0.16} ~~,~~~~
C_d = -0.40 \pm 0.41 \pm 0.06 ~,
\eeq
while Belle finds \cite{Nakahama:2007dg}
\beq
\mathcal{B}_d = (0.87^{+0.25}_{-0.20} \pm 0.09) \times 10^{-6} ~~,~~~~
S_d = -0.38^{+0.69}_{-0.77} \pm 0.09 ~~,~~~~
C_d = 0.38 \pm 0.38 \pm 0.05 ~.
\eeq
We see that essentially all values of $\sqrt{C_d^2 + S_d^2}$ are still
experimentally allowed.

Here is an example of the calculation of $|P_{uc}|$ using
Eq.~(\ref{eq:Puc2}).  We take the QCDf-inspired central value of CPS
for the branching ratio ($\mathcal{B}_d = 5 \times
10^{-7}$)\footnote{In fact, the branching ratio for $\bd \to K^{*0}
  {\bar K}^{*0}$ has been measured \cite{BdK*K*bar}. The world average
  is $\mathcal{B}_d = (8.1 \pm 2.3) \times 10^{-7}$ \cite{hfag}. In
  order to make the generalization of the CPS method more direct, in
  our analysis we use the CPS value for $\mathcal{B}_d$ (which differs
  from the experimental value by only a little more than $1\sigma$).},
and also take $0 \le \sqrt{C_d^2 + S_d^2} \le 1$. We compute
$|\lambda^{(d)}_u|$ using values for the various quantities taken from
the Particle Data Group \cite{pdg}.  Including the errors on these
quantities, we find that $|P_{uc}|$ can be as large as $1460 \pm 170$
eV ($Z_R$ positive) or $2060 \pm 240$ ($Z_R$ negative).  For
comparison, $|P_{uc}|$ is only about 180 eV if the CP asymmetries are
also fixed at the QCDf-inspired central values of CPS.  Note that the
discrete ambiguity with $Z_R$ negative corresponds to the case for
which $\bd$ decays are dominated by $P_{uc}$. On the other hand, we
naively expect $P_{tc}$ to be larger. Still, even if this solution
were discarded, the results of our analysis below would not be changed
fundamentally. The bottom line is that $|P_{uc}|$ can, in fact, be
large in $\bdkk$ decays.

We now return to $\bskk$ decays.  Even if $|P_{uc}|$ is large in $\bd$
decays, because of the $|\lambda^{(s)}_u|$ CKM suppression it is not
clear whether or not $|\lambda^{(s)}_u P'_{uc}|$ really plays a
significant role in the $\bs$ decays. In order to ascertain this, we
proceed as follows.  We apply the CPS method, but consider all
possible values of the observables in both $\bd$ and $\bs$
decays\footnote{In fact, the branching ratio for $\bs \to K^{*0} {\bar
    K}^{*0}$ has been measured \cite{BsK*K*bar}. Its value is $(2.81
  \pm 0.46~({\rm stat}) \pm 0.45~({\rm syst}) \pm 0.34~(f_s/f_d))
  \times 10^{-5}$. The CPS value for $\mathcal{B}_s$, which we use in
  our analysis, is $1.18 \times 10^{-5}$.}.  Thus, we use flavor SU(3)
symmetry to relate $|P_{uc}|$ and $|P'_{uc}|$, allowing for a 100\%
symmetry breaking. In order to study the worst-case scenario (the
largest possible value of $|P'_{uc}|$ within 100\% breaking), we fix
$|P'_{uc}| = 2 |P_{uc}|$. Thus, for example, for the case where the
branching ratio $\mathcal{B}_d$ is taken to be the QCDf-inspired
central value of CPS, but the CP asymmetries take all possible values,
$|P'_{uc}|$ can be as large as 2920 eV ($Z_R$ positive) or 4120 eV
($Z_R$ negative).

Given the worst-case value of $|P'_{uc}|$, assuming the CKM phases to
be known, and fixing $\beta_s = 0$ (in order to study the worst-case
prediction in the SM), only two parameters are left unknown in the
$\bskk$ decay.  These can be extracted from the branching ratio
$\mathcal{B}_s$ and the direct CP asymmetry $C_s$ (up to discrete
ambiguities, but this does not affect the following discussion).  Once
this is done, all the theoretical parameters in $\bskk$ are known, and
we can compute the time-dependent CP asymmetry $S_s$ and the effective
phase $\beta_s^{eff}$ ($\arg{(\bar\mathcal{A}_s / \mathcal{A}_s)}$).
Thus we get an evaluation of the (worst-case) theoretical uncertainty
of $\beta_s$ as extracted from the time-dependent CP asymmetry of
$\bskk$ decays.

\begin{figure}[htb]
	\centering
		\includegraphics[height=7.2cm]{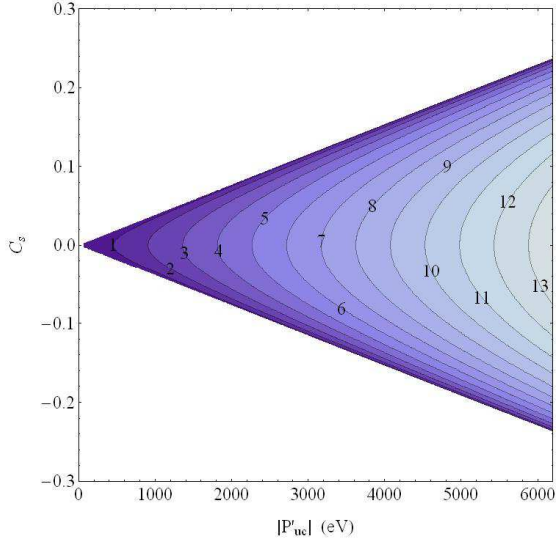}
\caption{Worst-case values of $\beta_s^{eff}$ (in degrees) as a
  function of $|P'_{uc}|$ and the direct CP asymmetry $C_s$.  The
  branching ratios are fixed to $\mathcal{B}_d = 5 \times 10^{-7}$ and
  $\mathcal{B}_s = 11.8 \times 10^{-6}$ (central values of CPS).}
\label{fig1}
\end{figure}

\begin{figure}
	\centering
		\includegraphics[height=7.2cm]{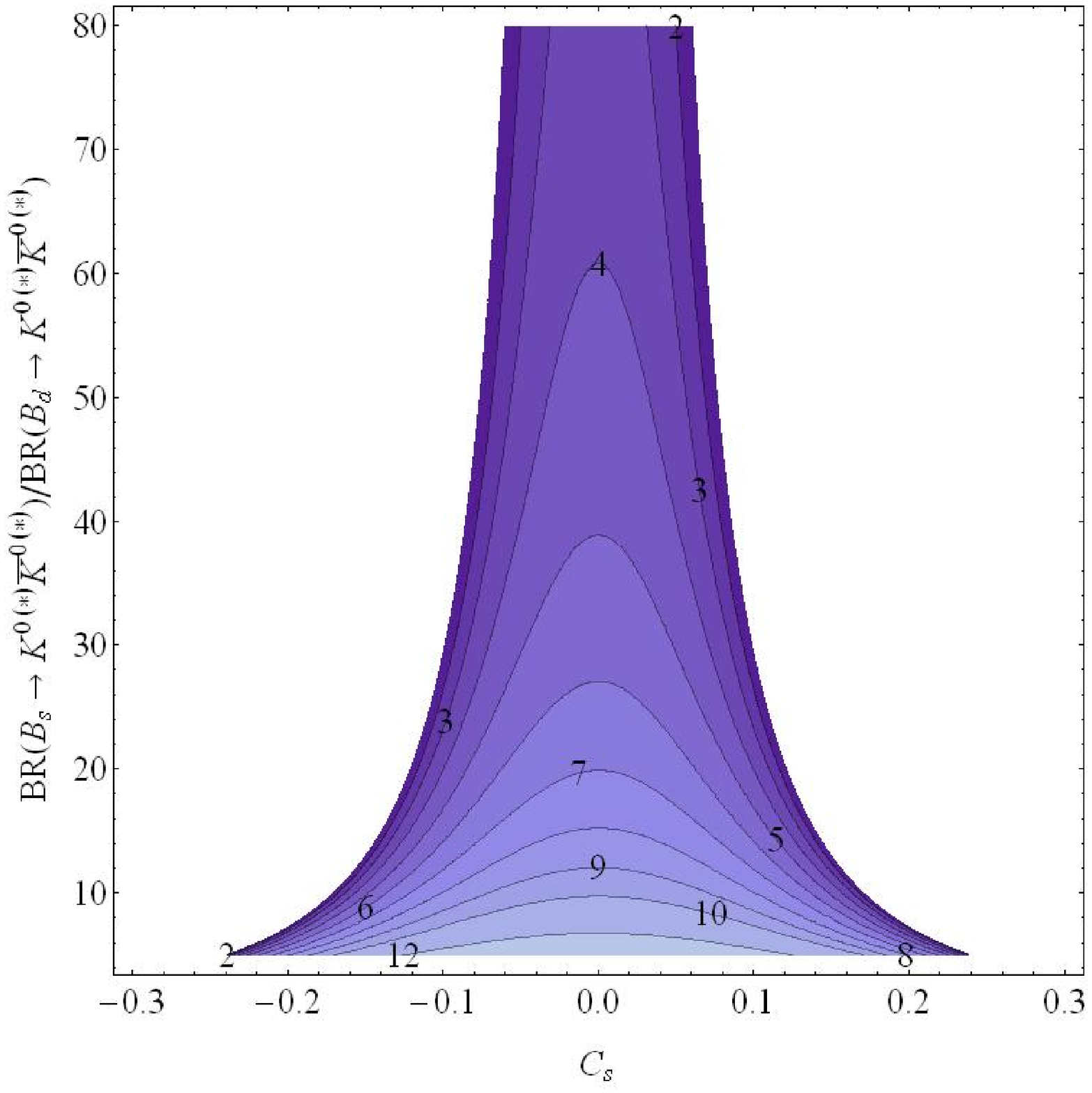}
		\includegraphics[height=7.2cm]{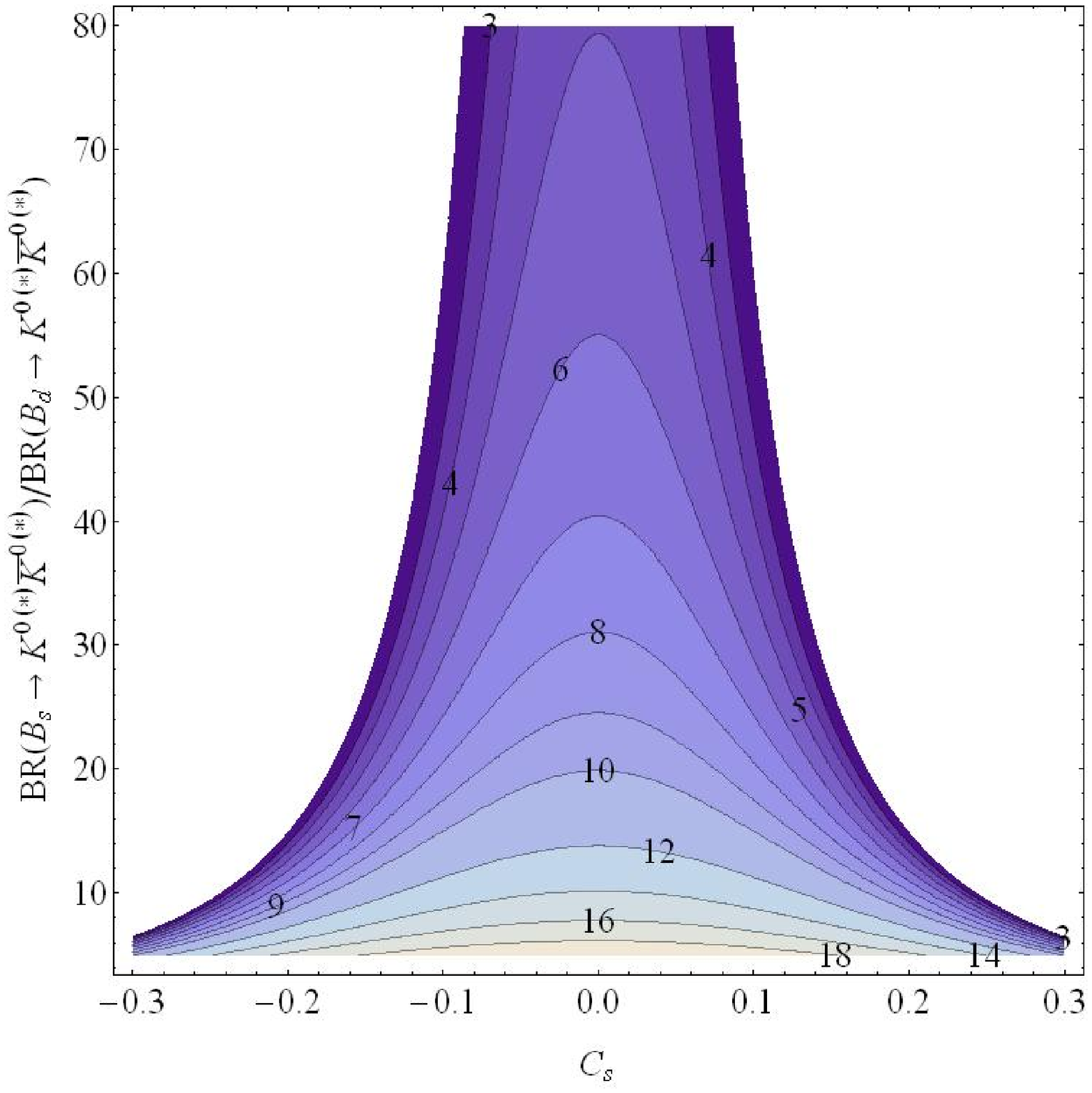}
\caption{Worst-case values of $\beta_s^{eff}$ (in degrees) as a
  function of the direct CP asymmetry $C_s$ and the ratio of branching
  ratios ($\mathcal{B}_s/\mathcal{B}_d$).  The plot on the left
  (right) is for $Z_R$ positive (negative) in Eq.~(\ref{eq:ZR}).}
\label{fig2}
\end{figure}

\begin{figure}
	\centering
		\includegraphics[height=7.2cm]{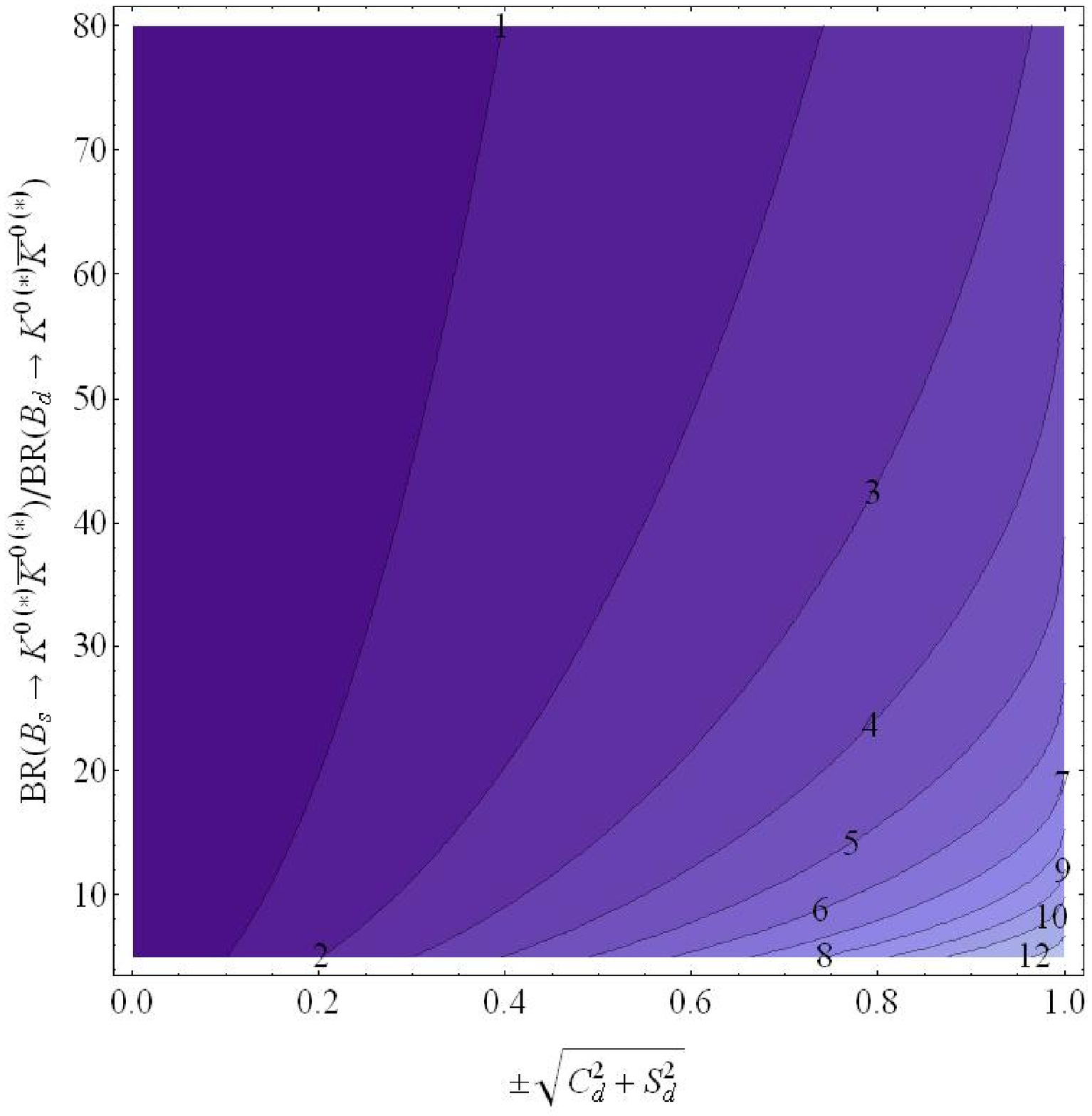}
		\includegraphics[height=7.2cm]{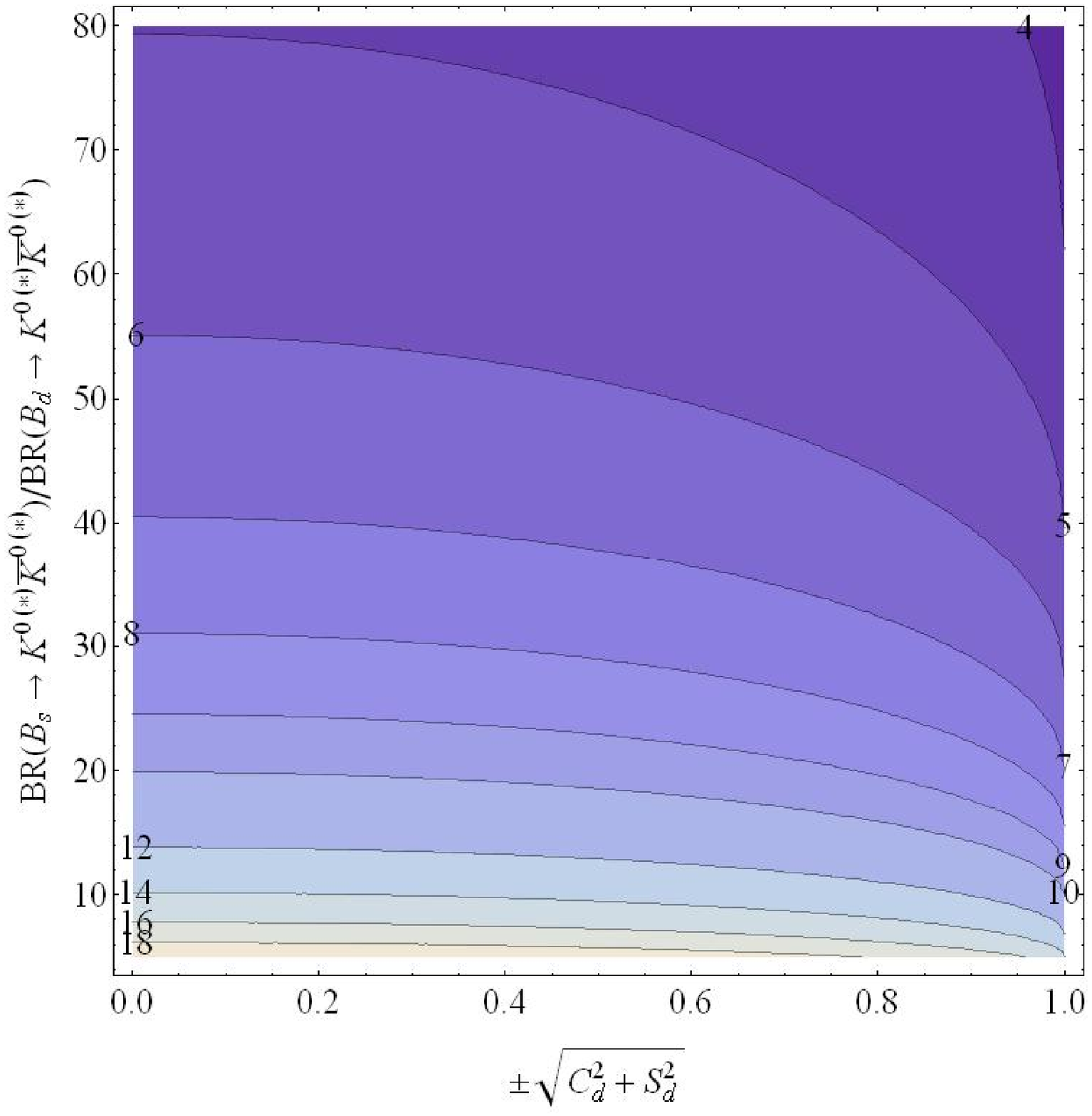}
\caption{Worst-case values of $\beta_s^{eff}$ (in degrees) as a
  function of $\sqrt{C_d^2+S_d^2}$ and the ratio of branching ratios
  ($\mathcal{B}_s/\mathcal{B}_d$).  The plot on the left (right) is
  for $Z_R$ positive (negative) in Eq.~(\ref{eq:ZR}).}
\label{fig3}
\end{figure}

We now present figures showing the worst-case $\beta_s^{eff}$ in
various situations.  The aim is to scan over the whole observable
space in order to ascertain how large $\beta_s^{eff}$ can be within
the SM.  In Fig.~\ref{fig1}, we fix both branching ratios to the CPS
central values, and give the worst-case value of $\beta_s^{eff}$ as a
function of $|P'_{uc}|$ and the direct CP asymmetry $C_s$.  The
effective phase is roughly proportional to $|P'_{uc}|$ and can be up
to $10^\circ$ in this restricted scenario. In Fig.~\ref{fig2}, we
repeat the calculation but also allow the branching ratios to vary,
presenting $\beta_s^{eff}$ as a function of $C_s$ and the ratio of
branching ratios ($\mathcal{B}_s/\mathcal{B}_d$).  In this case, for
the central maximum value of $|P_{uc}|$, an effective phase of up to
$12^\circ$ ($Z_R$ positive) or $18^\circ$ ($Z_R$ negative) is
obtained.  In Fig.~\ref{fig3}, $\beta_s^{eff}$ is given as a function
of $\sqrt{C_d^2+S_d^2}$ and $\mathcal{B}_s/\mathcal{B}_d$.

{}From the above figures, it is clear that $\beta_s^{eff}$ can be
large within the SM, and that the conclusions of CPS hold only for
certain sets of values of the experimental inputs.  Still, it is
interesting to note that a small theoretical error (say $\beta_s^{eff}
\le 5^\circ$) is found for a non-negligible subset of the input
numbers.  The general behavior of solutions is as follows:
\begin{enumerate}

\item for $Z_R$ positive, $\beta_s^{eff}$ is smaller for smaller
  values of $\sqrt{C_d^2+S_d^2}$ (it's the opposite for $Z_R$
  negative),

\item $\beta_s^{eff}$ is smaller for larger values of $|C_s|$ for
  fixed $|P'_{uc}|$,

\item $\beta_s^{eff}$ is smaller for larger values of
  $\mathcal{B}_s/\mathcal{B}_d$,

\item $\beta_s^{eff}$ is smaller for smaller values of SU(3) breaking.

\end{enumerate}
For the first three points we cannot do anything -- the measurements
of the observables are what they are.  The fourth point can be
understood as follows. The theoretical error $\beta_s^{eff}$ is due to
the presence of a nonzero $P'_{uc}$ in ${\cal A}_s$
[Eq.~(\ref{Bsamp})]. This error is roughly proportional to
$|P'_{uc}|$, which is itself equal to the product of $|P_{uc}|$ and an
SU(3)-breaking factor. For a given value of $|P_{uc}|$,
$\beta_s^{eff}$ is smaller if the SU(3)-breaking factor is
smaller. Thus, the assumption of CPS of 100\% breaking often leads to
a large $\beta_s^{eff}$.  The precise knowledge of the SU(3) breaking
between $|P_{uc}|$ and $|P'_{uc}|$ would therefore considerably reduce
the theoretical uncertainty on the extracted value of $\beta_s$ using
this method. The determination of the size of SU(3) breaking is
discussed in the next section.

\section{SU(3) Breaking}

As we have seen, the idea of obtaining information on $|P'_{uc}|$ by
relating it to $|P_{uc}|$ using flavor SU(3) is tenable. However, if
one simply takes an SU(3)-breaking factor of 100\%, this can lead to a
theoretical error on the extraction of $\beta_s$ of up to $18^\circ$.
Thus, in order to use this method, a better determination of the size
of SU(3) breaking must be found. In this section, we discuss ways,
both experimental and theoretical, of getting this information.

\subsection{Experimental Measurement of SU(3) Breaking}

\subsubsection{\boldmath $B^0_{d,s} \to K^{*0} {\bar K}^{*0}$}
\label{BKKSU3break}

The decays $B^0_{d,s} \to K^{(*)0} {\bar K}^{(*)0}$ really represent
three types of decay -- the final state can consist of $PP$, $PV$ or
$VV$ mesons ($P$ is pseudoscalar, $V$ is vector). Now, the CPS method
applies when the final state is a CP eigenstate. For $PP$ and $VV$
decays, this holds. However, $PV$ decays do not satisfy this
condition. Still, these decays can be used if the $K^{*0}$/${\bar
  K}^{*0}$ decays neutrally. That is, we have
\bea
B^0 & \to & \frac{1}{\sqrt{2}} \left( K^0 {\bar K}^{*0} + K^{*0} {\bar K}^0 \right) ~~~~~~ {\hbox{(CP eigenstate)}} \nn\\
& \to & K^0 {\bar K}^0 \pi^0 ~.
\label{PVdecay}
\eea
On the other hand, the CPS method cannot be used if the
$K^{*0}$/${\bar K}^{*0}$ decays to charged particles. This is because,
in this case, one cannot extract $|P_{uc}|$ from the $\bd$ decay --
there are more theoretical unknowns than observables.

The SM value of SU(3) breaking can be found from any single pair of
decays -- $|P_{uc}|$ and $|P'_{uc}|$ can be extracted from the $\bd$
and $\bs$ decays, respectively. (As we are interested in SU(3)
breaking in the SM, we set $\beta_s$ to zero.) In principle, this
value of SU(3) breaking ($|P'_{uc}|/|P_{uc}|$) can then be used in a
different decay, and the method of the previous section can be
applied. The problem here is that this approach is applicable only if
the SU(3) breaking in the two decays is expected to be
similar. However, $PP$, $PV$ and $VV$ decays are all different
dynamically, so that there is no a-priori reason to expect this to
hold. For example, the decay of Eq.~(\ref{PVdecay}) is very different
from the $PP$ decay $B^0 \to K^0 {\bar K}^0$, and so the $PV$ and $PP$
SU(3) breakings are not likely to be similar.

There is one exception, and it involves the $VV$ decays $B^0_{d,s} \to
K^{*0} {\bar K}^{*0}$. Since the final-state particles are vector
mesons, when the spin of these particles is taken into account, these
decays are in fact three separate decays, one for each polarization.
The polarizations are either longitudinal ($A_0$), or transverse to
their directions of motion and parallel ($A_\|$) or perpendicular
($A_\perp$) to one another. By performing an angular analysis of these
decays, the three polarization pieces can be separated. 

It is also possible to express the polarization amplitudes using the
helicity formalism. Here, the transverse amplitudes are written as
\bea
A_\| &=& \frac{1}{\sqrt{2}} (A_+ + A_-) ~, \nn\\
A_\perp &=& \frac{1}{\sqrt{2}} (A_+ - A_-) ~.
\eea
However, in the SM, the helicity amplitudes obey the hierarchy
\cite{BRY,Kagan}
\beq
\left\vert \frac{A_+}{A_-} \right\vert = \frac{\Lambda_{QCD}}{m_b} ~.
\eeq
That is, in the heavy-quark limit, $A_+$ is negligible compared to
$A_-$, so that $A_\| = -A_\perp$. Thus, one expects the SU(3) breaking
for the $\|$ and $\perp$ polarizations to be approximately equal.
One can therefore extract $|P_{uc}|$ and $|P'_{uc}|$ from the $\bd$
and $\bs$ decays for one of the transverse polarizations, compute the
SU(3) breaking ($|P'_{uc}|/|P_{uc}|$), and apply this value of SU(3)
breaking to the other transverse polarization decay pair.  In this way
the SU(3)-breaking factor can be measured experimentally, and can be
used to determine the theoretical uncertainty in the extraction of
$\beta_s$.

\subsubsection{\boldmath $B^+ \to K^+ \bar K^0$ and $B^+ \to \pi^+ K^0$}

Other decays which can be used to measure SU(3) breaking are the
U-spin-conjugate pair $B^+ \to K^+ \bar K^0$ and $B^+ \to \pi^+
K^0$. While it is true that these do not involve $\bs$ and $\bd$
mesons, both are pure penguin decays, just like $B^0_{d,s} \to
K^{(*)0} {\bar K}^{(*)0}$.  Restricting ourselves to the $PP$ final
states, we then expect that the SU(3) breaking in $B^+ \to K^+ \bar
K^0$ and $B^+ \to \pi^+ K^0$ is similar (though not necessarily equal)
to that in $\bd \to K^0 {\bar K}^0$ and $\bs \to K^0 {\bar K}^0$.
The measurement of SU(3) breaking can therefore be done using the
$B^+$ decays and applied to the $\bd$/$\bs$ decays.

There is a difference compared to the previous example. Since there
are no indirect CP asymmetries in $B^+$ decays, one cannot measure
$|P'_{uc}|/|P_{uc}|$. The SU(3) breaking probed in $B^+ \to K^+ \bar
K^0$ and $B^+ \to \pi^+ K^0$ is
\bea
\label{Y'/Y}
-\frac{Y'}{Y} & = &
\frac{|\lambda^{(s)}_u||\lambda^{(s)}_t|}{|\lambda^{(d)}_u||\lambda^{(d)}_t|} \,
\frac{\sin{\gamma}}{\sin{\alpha}} \, 
\frac{|P'_{uc}|}{|P_{uc}|} \, \frac{|P'_{tc}|}{|P_{tc}|} \, \frac{\sin\delta'}{\sin\delta} \nn\\
& = &
\frac{|P'_{uc}|}{|P_{uc}|} \, \frac{|P'_{tc}|}{|P_{tc}|} \, \frac{\sin\delta'}{\sin\delta} ~.
\eea
In the second line, all the CKM factors cancel due to the sine law
associated with the unitarity triangle.  Thus, if the $B^+$ decay pair
is used to measure the SU(3) breaking, the theoretical error in the
extraction of $\beta_s$ must be calculated relating $|P'_{uc}|
|P'_{tc}| \sin\delta'$ of $\bs \to K^0 {\bar K}^0$ to $|P_{uc}|
|P_{tc}| \sin\delta$ of $\bd \to K^0 {\bar K}^0$.

\subsubsection{Other SU(3) pairs}

There are many other pairs of decays that are related by U spin or
SU(3): $\bd \to \pi^+ \pi^-$ and $\bs \to K^+ K^-$, $\bd \to \pi^0K^0$
and $\bs \to \pi^0 \kbar$, etc. A complete list of two- and three-body
decay pairs, as well as a discussion of the measurement of
U-spin/SU(3) breaking, is given in Ref.~\cite{SU3break}.  For some of
them we already have measurements of the breaking.

For example, consider the pair $\bs \to \pi^+ K^-$ and $\bd \to \pi^-
K^+$. The measurement of the SU(3) breaking of Eq.~(\ref{Y'/Y}) gives
\cite{SU3break}
\beq
-\frac{Y'}{Y} = 0.92 \pm 0.42 ~.
\eeq
Although the error is still substantial, we see that the central value
implies small SU(3) breaking.  The problem is that $\bs \to \pi^+ K^-$
and $\bd \to \pi^- K^+$ are not pure-penguin decays, so that it is not
clear how the above SU(3) breaking is related to that in $B^0_{d,s}
\to K^{(*)0} {\bar K}^{(*)0}$, if at all.  Still, if one measures the
SU(3) breaking in several different decay pairs, it can give us a
rough indication as to what to take for $B^0_{d,s} \to K^{(*)0} {\bar
  K}^{(*)0}$.

\subsection{Theoretical Input on SU(3) Breaking}

Consider again the $\bskk$ amplitude, Eq.~(\ref{Bsamp}). If the
$t$-quark contribution is eliminated using the unitarity of the CKM
matrix, we have
\beq
{\cal A}_s = T' \lambda^{(s)}_u + P' \lambda^{(s)}_c ~,
\eeq
where $T' \equiv P'_u - P'_t$, $P' \equiv P'_c - P'_t$.  Now, in QCDf
$T'$ and $P'$ are calculated using a systematic expansion in
$1/m_b$. However, a potential problem occurs because the higher-order
power-suppressed hadronic effects contain some chirally-enhanced
infrared (IR) divergences. In order to calculate these, one introduces
an arbitrary infrared (IR) cutoff. The key observation here is that
the difference $T' - P'$ is free of these dangerous IR divergences
\cite{DMV}.  And, although the calculation of various hadronic
quantities in pQCD is different than in QCDf, the difference $T'-P'$
is the same in both formulations. This also holds for $T - P$ in the
$\bdkk$ amplitude.

Since $T' - P' = P'_{uc}$ and $T - P = P_{uc}$, this suggests that the
calculation of $|P'_{uc}|$ and $|P_{uc}|$ is under good control
theoretically. This allows us to calculate $|P'_{uc}|/|P_{uc}|$, which
gives us the theoretical prediction of SU(3) breaking. There are many
quantities which enter into the calculation of $|P'_{uc}|$ and
$|P_{uc}|$ -- the renormalization scale $\mu$, the Gegenbauer
coefficients in the light-cone distributions, the quark masses,
etc.\ -- and the errors on these quantities are quite large at
present. However, most of these quantities and their errors cancel in
the ratio $|P'_{uc}|/|P_{uc}|$. For the various $B^0_{d,s} \to
K^{(*)0} {\bar K}^{(*)0}$ decays we find
\bea
\label{SU3break}
PP &:& \frac{|P'_{uc}|}{|P_{uc}|} =
\frac{M_{B_s}^2 F_0^{\bs \to K}(M_K^2)}{M_{B_d}^2 F_0^{\bd \to K}(M_K^2)} = 0.86 \pm 0.15 ~,\nn\\
PV &:& \frac{|P'_{uc}|}{|P_{uc}|} = \frac{M_{B_s}^2 F_+^{\bs \to K}(M_{K^*}^{2})}{M_{B_d}^2 F_+^{\bd \to K}(M_{K^*}^{2})} = 0.86 \pm 0.15 ~,\nn\\
VP, VV_0 &:& \frac{|P'_{uc}|}{|P_{uc}|} = \frac{M_{B_s}^2 A_0^{\bs \to K^*}(M_{K^{(*)}}^{2})}{M_{B_d}^2 A_0^{\bd \to K^*}(M_{K^{(*)}}^{2})} = 0.87 \pm 0.19 ~,\\
VV_{\|}, VV_{\perp} &:& \frac{|P'_{uc}|}{|P_{uc}|} =  \frac{M_{B_s} ( F_-^{\bs \to K^*}(M_{K^*}^{2}) \pm F_+^{\bs \to K^*}(M_{K^*}^{2}))}{M_{B_d} (F_-^{\bd \to K^*}(M_{K^*}^{2}) \pm F_+^{\bd \to K^*}(M_{K^*}^{2}))} = 0.79 \pm 0.16~. \nn
\eea

Above we have taken $F^{B \to K^{(*)}}(M_{K^{(*)}}^2) \simeq F^{B\to
  K^{(*)}}(0)$ since the variation of the ratio of form factors over
this range of $q^2$ falls well within the errors of their calculation
\cite{BRY, LCSR}. For $PV$ and $VP$ decays, the spectator quark goes
in the first meson. In the last expression, we have $F_+^{B \to K^*} =
0.00 \pm 0.06$ \cite{BRY}, so that one has the same SU(3) breaking for
the $\|$ and $\perp$ polarizations. As discussed in
Sec.~\ref{BKKSU3break}, this is to be expected.

Now, the QCDf calculation is to $O(\alpha_s)$, and the above
expression indicates that, to this order, the SU(3)-breaking term is
factorizable. Thus, the theoretical prediction is fairly robust.  On
the other hand, SCET says that there are long-distance contributions
to $P'_c$ and $P_c$. Although this could introduce some uncertainty
into $|P'_{uc}|/|P_{uc}|$, there might also be a partial cancellation
in the ratio. Our point here is that, though one generally wants to
avoid theoretical input, since this is largely based on models, the
SU(3) breaking in $|P'_{uc}|/|P_{uc}|$ may be theoretically clean.

In Sec.~\ref{BKKSU3break}, it was noted that the CPS method can be
used when the final state in $B^0_{d,s} \to K^{(*)0} {\bar K}^{(*)0}$
is a CP eigenstate. Thus, if one wishes to use the theoretical input
of Eq.~(\ref{SU3break}), one can simply apply it to $PP$ or $VV$
decays. However, this does not hold for $PV$ or $VP$ decays, which are
not CP eigenstates. Still, one can use the CPS method on the decay of
Eq.~(\ref{PVdecay}), which is a linear combination of $PV$ and $VP$
states. And, since the theoretical $PV$ SU(3) breaking in
Eq.~(\ref{SU3break}) is about equal to that of $VP$, this theoretical
input can be applied to the $PV + VP$ decay.

Finally, in Sec.~\ref{bskkSec} we noted that, in the presence of a
nonzero $P'_{uc}$, one cannot cleanly extract $\beta_s$ from $\bskk$
-- one needs additional input. In Ref.~\cite{DMV2} it was the above
theoretical calculation of $|P'_{uc}|$ which was taken as the
input. We note that the method using $\bskk$ and $\bdkk$ is somewhat
more precise since most of the errors in the calculation of
$|P'_{uc}|$ cancel in the SU(3)-breaking ratio of
Eq.~(\ref{SU3break}).

\section{Conclusions}

The pure $\btos$ penguin decay $\bskk$ is potentially a good candidate
for measuring the $\bs$-$\bsbar$ mixing phase, $\beta_s$. If its
amplitude were dominated by $V_{tb}^* V_{ts} P'_{tc}$, the indirect CP
asymmetry would simply measure $\beta_s$. Unfortunately, although the
second contributing amplitude, $V_{ub}^* V_{us} P'_{uc}$, is expected
to be small, it is not clear that it is completely negligible. A
nonzero $V_{ub}^* V_{us} P'_{uc}$ can change the extracted value of
$\beta_s$ from its true value, i.e.\ it can lead to a theoretical
error. Since the measurement of $\beta_s$ is an important step in the
search for new physics, the size of this theoretical error is
important.

The size of $P'_{uc}$ has been examined by Ciuchini, Pierini and
Silvestrini (CPS). They note that the amplitude $P_{uc}$ can be
extracted from the U-spin-conjugate decay, $\bdkk$, and can be related
to $P'_{uc}$ by SU(3). They choose values for the $\bdkk$ experimental
observables inspired by QCDf, allow for 100\% SU(3) breaking, and
compute $P'_{uc}$. They find that the theoretical error on $\beta_s$
is very small, i.e.\ that the presence of the $V_{ub}^* V_{us}
P'_{uc}$ amplitude has little effect on the extraction of $\beta_s$.

In this paper, we revisit the CPS method. In particular, we consider
most values of the $\bdkk$ observables, still allowing for 100\% SU(3)
breaking. We find that, although the theoretical error remains small
for a significant subset of these input values, it can be large for
other values. We find that an error of up to $18^\circ$ is possible,
which makes the extraction of $\beta_s$ from $\bskk$ problematic.

This issue can be resolved if we knew the value of SU(3) breaking. We
therefore discuss different ways, both experimental and theoretical,
of determining this quantity. From the experimental point of view, the
size of SU(3) breaking can be measured using a different $\bd$/$\bs$
decay pair. We show that the $VV$ decay $B^0_{d,s} \to K^{*0} {\bar
  K}^{*0}$ or $B^+ \to K^+ \bar K^0$/$B^+ \to \pi^+ K^0$ can be used
in this regard. It is also possible to use theoretical input. Within
QCDf, the SU(3)-breaking term is factorizable, and so the theoretical
prediction for this quantity may be reasonably clean.

\bigskip
\noindent
{\bf Acknowledgments}: We thank Tim Gershon for helpful
communications. This work was financially supported by NSERC of Canada
(BB, DL), by the US-Egypt Joint Board on Scientific and Technological
Co-operation award (Project ID: 1855) administered by the US
Department of Agriculture and in part by the National Science
Foundation under Grant No.\ NSF PHY-1068052 (AD), and by FQRNT of
Qu\'ebec (MI).


\end{document}